\documentclass[letterpaper]{article} 
\usepackage[preprint]{aaai2027}  
\usepackage[hyphens]{url}  
\usepackage{graphicx} 
\usepackage[numbers,sort&compress]{natbib} 
\usepackage{caption} 
\usepackage{booktabs}

\usepackage{multirow}
\usepackage{xcolor}

\title{SWE-Gate: Passing Functional Tests Is Not Enough for Software Engineering Agents }
\author{
    Xin He\textsuperscript{\rm 1},
   Yanlin Wang\textsuperscript{\rm 1}\thanks{Yanlin Wang is the corresponding author.},
    Mingwei Liu\textsuperscript{\rm 1},
    Jiachi Chen\textsuperscript{\rm 2},
    Hongyu Zhang\textsuperscript{\rm 3},
    Guanbin Li\textsuperscript{\rm 4}
}

\affiliations{
    \textsuperscript{\rm 1}School of Software Engineering, Sun Yat-sen University  \\
    \textsuperscript{\rm 2}College of Computer Science and Technology, Zhejiang University \\
    \textsuperscript{\rm 3}School of Big Data and Software Engineering, Chongqing University \\
    \textsuperscript{\rm 4}School of Computer Science and Engineering, Sun Yat-sen University \\
}

\begin{document}

\maketitle

\begin{abstract}
Repository-level software engineering benchmarks have significantly advanced the evaluation of coding agents, but existing benchmarks primarily measure whether generated patches pass functional tests and overlook review-derived acceptance constraints (review constraints) that often influence whether a patch is acceptable in real-world software development. We introduce SWE-Gate, a repository-level benchmark for software engineering agents that explicitly evaluates review constraint compliance alongside functional correctness. SWE-Gate derives review constraints from real pull request review comments and synthesizes repository-level repair instances around these constraints. Each instance provides separate functional and constraint tests, together with non-compliant and gold patches, enabling explicit separation between issue resolution capability and review constraint compliance. We construct SWE-Gate with 303 repository-level repair instances spanning 75 open-source Python repositories across diverse software domains. Experiments with four LLM backends spanning different capability levels under a common coding-agent scaffold reveal a substantial gap between functional success and success under the complete repair specification: among 644 repairs that pass the functional tests, 221 fail to satisfy the provided review constraints. These findings show that functional-only evaluation overestimates agents' ability to satisfy the full requirements of repository-level repair tasks. The replication package including code, data, and experimental results is available at \url{https://github.com/DeepSoftwareAnalytics/SWE-Gate}.
\end{abstract}

\section{Introduction}

Recent advances in large language models (LLMs) have transformed automated software engineering, evolving from early function-level code generation benchmarks \cite{shi2023sotanaopensourcesoftwaredevelopment,wang2024agentssoftwareengineeringsurvey} to repository-level agents capable of understanding large codebases \cite{zhang-etal-2023-repocoder,zhang2025llmhallucinationspracticalcode,zheng2025generalperformancedomain}, localizing defects, and repairing real software issues using code, version history, tests, and natural-language reports \cite{jimenez2024swebench,NEURIPS2024_5a7c9475,wang2025openhands,xia2024agentlessdemystifyingllmbasedsoftware}. Repository-level repair has consequently become an important measure of practical programming ability. SWE-Bench \cite{jimenez2024swebench}, the de facto benchmark, constructs tasks from real issue--pull request pairs and evaluates patches with executable functional tests, driving systems such as SWE-Agent \cite{NEURIPS2024_5a7c9475} and OpenHands \cite{wang2025openhands}. 

Despite their success, existing repository-level benchmarks largely adopt a single evaluation criterion: a repair is considered successful if the generated patch passes the functional test suite which is typically provided with the PR and mainly assesses whether the reported issue has been functionally resolved. However, in practice, a patch that resolves the reported issue and passes functional tests is not necessarily accepted \cite{10.1145/2771783.2771791,yu-etal-2025-utboost}. \textbf{During code review, repository maintainers frequently require contributors to satisfy additional requirements beyond functional correctness. We call such requirements \emph{review-derived acceptance constraints} (\emph{gates}): additional, objectively testable requirements that restrict which functionally correct repairs are acceptable.} For brevity, we refer to them as \textbf{\emph{review constraints}} hereafter. These requirements are often reflected in review comments that request changes before a patch can be accepted. Examples of such requirements include preserving backward compatibility, maintaining established exception semantics, or following repository-specific implementation conventions \cite{6606617,10.1145/3183519.3183525,10.1145/2491411.2491444}. 

For example, a Pydantic pull request sought to add conditional serialization for extra fields. Its initial implementation provided the requested behavior by introducing a new core-schema representation, but a reviewer noted that changing this public interface would be breaking and recommended adding an \texttt{extras\_ser\_exclude\_if} parameter instead.\footnote{\url{https://github.com/pydantic/pydantic/pull/12657#discussion_r2671466868}} The contributor subsequently adopted this backward-compatible design.

The two requirements can be tested separately. A functional test checks whether qualifying extra fields are omitted from serialized output, whereas a compatibility test checks whether the existing public schema remains unchanged and usable by code written against the previous format. A patch may therefore pass the functional test while failing compatibility, showing that implementing the requested behavior does not necessarily satisfy the review constraint raised during review.

Existing benchmarks answer whether a model can generate a functionally correct repair, but not whether that repair also complies with review-derived review constraints. These are separable evaluation dimensions: issue tests establish functional success, whereas constraint tests determine whether a functional repair also meets additional acceptance requirements. Evaluating only the first may therefore overestimate agents' ability to satisfy complete repair requirements.

To bridge this gap, we introduce \textbf{SWE-Gate}, the first repository-level benchmark to evaluate review constraint compliance alongside functional correctness using separate executable tests. Rather than manually inventing rules or attaching them to unrelated tasks, SWE-Gate derives natural-language constraints from maintainer reviews and synthesizes realistic repairs around their engineering intent. Each instance includes a non-compliant patch that passes functional tests but violates the constraint and a gold patch that passes both, demonstrating that the dimensions are separable, executable, and jointly satisfiable.

Using SWE-Gate, we evaluate representative LLMs on 303 repository-level repair instances spanning 75 repositories across six software domains. Our experiments reveal that many repairs that successfully pass functional tests nevertheless violate review constraints, demonstrating that functional success alone does not establish that an agent-generated repair satisfies the full set of repository requirements or is acceptable for integration.

Our contributions are summarized as follows:

\begin{itemize}

\item We introduce \textbf{SWE-Gate}, the first repository-level software engineering benchmark that incorporates executable review constraints into realistic software repair tasks.

\item We propose a dual-dimension evaluation protocol that separately measures functional correctness and review constraint compliance with executable tests.

\item We develop a semi-automated construction framework and use it to create 303 quality-assured instances across 75 repositories and six domains, then characterize constraint following across four LLM backends evaluated under a common Mini-SWE-Agent scaffold.

\end{itemize}

\section{Related Work}
Early benchmarks such as HumanEval and MBPP evaluate function-level programs with unit tests \cite{chen2021evaluatinglargelanguagemodels,austin2021programsynthesislargelanguage}, whereas RepoBench, RepoCoder, and CrossCodeEval introduce repository or cross-file context, with subsequent work further studying repository-level code generation\cite{liu2023repobenchbenchmarkingrepositorylevelcode,zhang-etal-2023-repocoder,ding2023crosscodeeval,li2024repomincoder}; multilingual code evaluation and real-fault corpora are further represented by xCodeEval, Defects4J, and BugsInPy \cite{khan2023xcodeevallargescalemultilingual,10.1145/2610384.2628055,10.1145/3368089.3417943}. Pre-trained code models have substantially advanced code representation and understanding \cite{guo2022unixcoder,lin2026bettercodeunderstanding}, while subsequent work has incorporated richer contextual information into code intelligence\cite{guo2023snippetcomment,wang2024sparsecoder,wang2021cocosum}. Recent studies further examine context utilization, task-relevant retrieval, and the quality of repository-level agent trajectories, including work on context use and retrieval \cite{wang2026contextutilization,wang2025rlcoder,jiang2026aligncoderaligningretrievaltarget,wang2026reporeasoner,gu2026retrievalcodegen} and on selecting high-quality trajectories for more effective agent supervision \cite{zheng2026sweprimefewertrajectoriesbetter}, advancing the study of LLMs for repository-level software engineering tasks.

SWE-bench formulates repository-level repair as resolving real GitHub issues with executable tests \cite{jimenez2024swebench}, motivating agents such as SWE-agent, OpenHands, and Agentless \cite{NEURIPS2024_5a7c9475,wang2025openhands,xia2024agentlessdemystifyingllmbasedsoftware}. Multi-agent approaches such as MAGIS further explore collaborative issue resolution, while recent repair-agent research also investigates how agents explore and select repair strategies \cite{magis2024,jiang2026phoenixrepair}, reflecting the broader adoption of LLM-based agents in software engineering. Later benchmarks extend languages, enterprise projects, repository evolution, freshness, contamination control, multimodal issues, and security  \cite{NEURIPS2025_5afa9cb1,zan2024swebenchjavagithubissueresolving,deng2025swebenchproaiagents,joshi2025swebenchclcontinuallearningcoding,NEURIPS2025_d83c4a74,NEURIPS2025_21bec6ac,yang2024swebenchmultimodalaisystems,guo2025omnigirl,zheng2025humanevo,wang2024repotransbench,arkrepobench2026,realsecbench2026}, but still primarily evaluate tested functional behavior. Related empirical studies have also highlighted that benchmark design and contextual assumptions can substantially affect the realism of LLM-based code-generation evaluation \cite{zheng2024realisticevaluation}.
Scalable task construction has been explored by SWE-smith, which synthesizes test-breaking tasks within repositories \cite{NEURIPS2025_8b86cf5a}, and SWE-Mirror, which transfers the semantic essence of real issues across repositories \cite{wang2025swemirrorscalingissueresolvingdatasets}. Recent work further investigates automated construction of software engineering datasets, including automated SWE data construction and LLM-assisted rebuilding of code-intelligence benchmarks \cite{Guo2025SWEDC,LargeLanguageModelsAreQualifiedBenchmarkBuilders}. SWE-Gate is inspired by SWE-Mirror's cross-repository transfer paradigm but instead transfers review-derived review constraints together with their bug patterns, applicability conditions, and validation requirements, producing separate functional and constraint tests as well as non-compliant and gold repairs.

Passing available tests does not necessarily establish patch correctness because weak suites can admit overfitting repairs \cite{10.1145/2771783.2771791,10.1145/2786805.2786825,10.1145/3106237.3106274}, a problem also observed in SWE-bench evaluations \cite{yu-etal-2025-utboost}. Moreover, code reviews address maintainability, consistency, compatibility, and other engineering concerns beyond functional defects \cite{6606617,10.1145/2597073.2597082,10.1145/3183519.3183525,10.1145/2491411.2491444}. These concerns also require engineering knowledge beyond the target code, with project and testing knowledge shown to improve LLM-based test generation \cite{li2026knowledgeunittest}. More broadly, surveys of code LLMs and LLM-based software engineering document the limitations of narrow evaluation criteria for complex development tasks \cite{zheng2024surveylargelanguagemodels,zheng2024understandinglargelanguagemodels}. Constraint-aware code benchmarks have begun evaluating additional requirements \cite{duan2025hierarchicalevolvablebenchmarkfinegrained}, while repository-level work attaches design constraints but relies on LLM-based verification \cite{Yu2026DoesPR}; such judges can exhibit subjectivity and evaluation biases \cite{NEURIPS2023_91f18a12,wang-etal-2024-large-language-models-fair}. In contrast, SWE-Gate constructs every instance around a review-derived constraint and separately executes functional and constraint tests, enabling deterministic evaluation of both dimensions.

\section{SWE-Gate Benchmark}

\begin{figure*}[t]
    \centering
    \includegraphics[width=\textwidth]{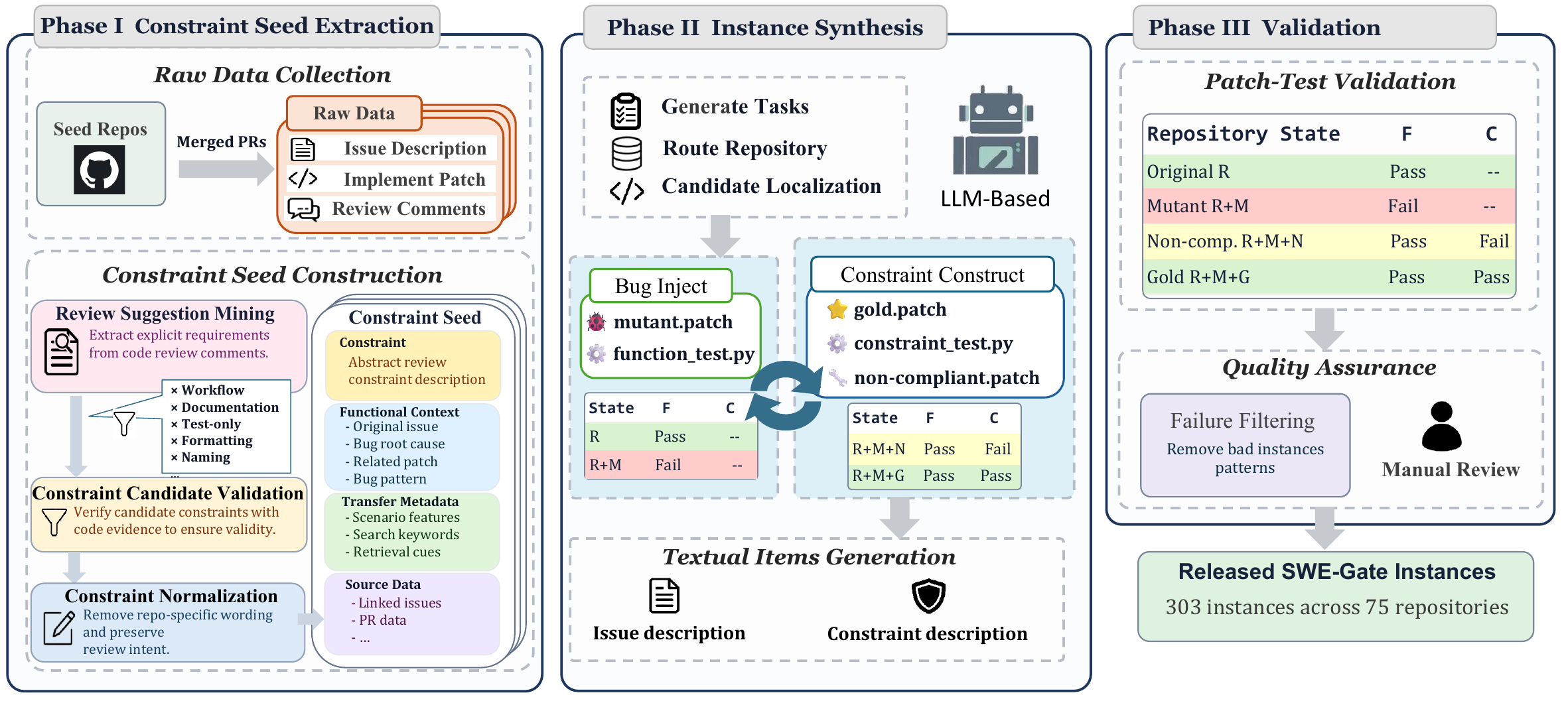}
    \caption{Overview of the SWE-Gate construction pipeline. SWE-Gate reconstructs issue--pull request artifacts, uses LLM-based processing to extract atomic review suggestions and select verifiable constraint seeds, transfers the seeds into compatible repository contexts, and validates the resulting instances through executable tests and quality assurance.}
    \label{fig:construction_pipeline}
\end{figure*}

SWE-Gate adopts a constraint-first construction strategy because a review-derived constraint is tied to a particular functional and code context rather than being a standalone rule that can be appended to an arbitrary issue. Moreover, directly reusing the source issue--pull request pair preserves a public one-to-one correspondence between the task and its known repair. SWE-Gate instead jointly abstracts the bug pattern, engineering intent, and applicability conditions from real review discussions and instantiates them in distinct, compatible repository contexts. This strategy preserves the authenticity of the original review requirement, expands a single constraint into multiple contextually valid tasks across repositories and domains, and mitigates direct solution memorization without claiming to eliminate all training-data contamination. Figure~\ref{fig:construction_pipeline} illustrates the resulting construction pipeline.

\subsection{Benchmark Instance}

\begin{figure}[t]
    \centering
    \includegraphics[width=\columnwidth]{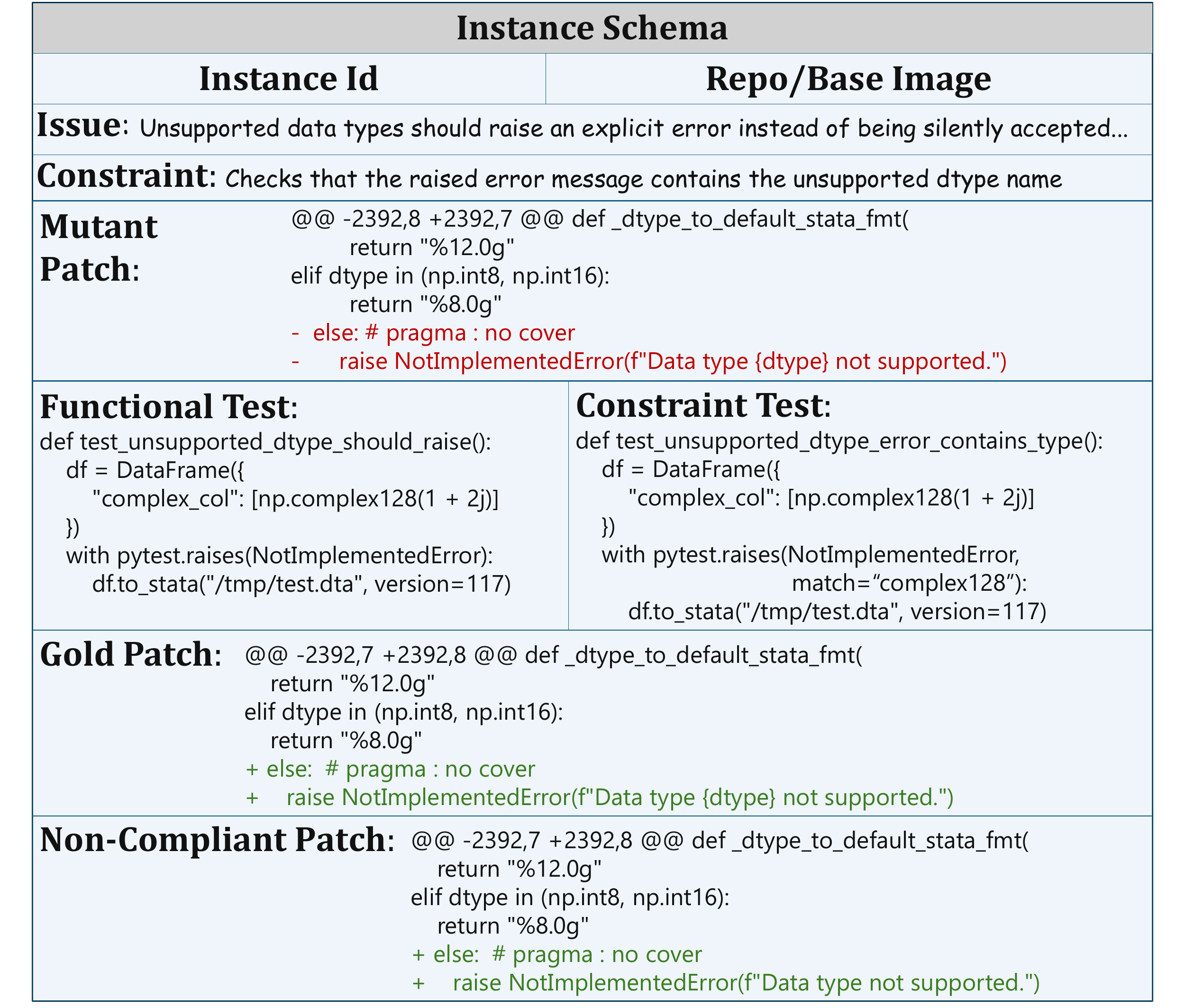}
    \caption{SWE-Gate benchmark instance schema.}
    \label{fig:instance_schema}
\end{figure}

Each SWE-Gate instance is a repository-level repair task that combines natural-language descriptions, patches, and executable tests to evaluate issue resolution and constraint following (Figure~\ref{fig:instance_schema}).

\begin{itemize}
    \item \textbf{Issue Description}. A natural-language specification of the observed incorrect behavior and expected functionality, without revealing the root cause or repair strategy.
    \item \textbf{Mutant Patch}. A patch that injects a reproducible target defect into the original repository while preserving the surrounding engineering context.
    \item \textbf{Functional Test}. An executable test that exposes the defect and validates its repair.
    \item \textbf{Constraint Description}. A natural-language engineering requirement that the repair must satisfy beyond functional correctness.
    \item \textbf{Constraint Test}. An executable test that validates compliance with the review constraint.
    \item \textbf{Non-compliant Patch}. A reference repair that passes the functional test but violates the constraint.
    \item \textbf{Gold Patch}. A reference repair that passes both the functional and constraint tests.
\end{itemize}

Among these artifacts, the non-compliant patch and the gold patch play a central role in the design of SWE-Gate instances. The non-compliant patch demonstrates that a patch can fix the issue while violating the review constraint; therefore, the review constraint is not merely a restatement of the functional requirement. The gold patch demonstrates that the review constraint is satisfiable and can hold simultaneously with the functional repair. Together, they show that constraint compliance is not implied by functional correctness, can be validated separately, and is jointly satisfiable with a functional repair.

\subsection{Constraint-First Instance Construction}

\subsubsection{Repository Selection}

SWE-Gate distinguishes between two types of repository roles: \emph{seed repositories} and \emph{instance repositories}. Seed repositories provide engineering knowledge from real code review discussions, while instance repositories provide the code contexts needed to construct executable benchmark instances.

Seed repositories are mature open-source Python projects. These projects typically have active development communities, rich pull request review histories, and high community adoption. We select such repositories because their review comments are more likely to reflect engineering requirements raised by maintainers during real software development.

Because constraints are context-dependent, candidate instance repositories are selected from related functionality or domains; for example, data-processing constraints prioritize that ecosystem and CLI constraints prioritize command-line projects. This domain mapping only defines a search space rather than presuming transferability: synthesis must still verify every target location against the seed's semantic and validation requirements. Representative mappings appear in the supplementary material . 

\subsubsection{Constraint Extraction}
Before the two LLM-assisted extraction stages, SWE-Gate collects merged pull requests from seed repositories through the GitHub API. For each pull request, we retrieve the pull request itself, review comments, linked issues, changed files, and the final unified diff. Inline review comments retain their file paths, line locations, and diff hunks. Together, these artifacts form the raw data used as input to the first stage.

Inspired by the atomic-suggestion stage of DesignHunter in SWE-Shield \cite{Yu2026DoesPR}, the first-stage LLM extracts only suggestions explicitly stated in review comments. It decomposes comments containing multiple requests into atomic records of the reviewer-identified problem, requested change, rationale, category, source identifiers, and confidence, without inferring unstated rules. Rule-based filtering then removes workflow, documentation-only, test-only, formatting, naming, typographical, vague, and otherwise unverifiable requests. Linking each retained record to its source comment and original and revised diff context preserves review and implementation traceability.

Second, deterministic checks require a linked issue, substantive suggestion, and sufficient diff context. An LLM then evaluates each survivor with its issue, review, and code evidence. Candidates are retained only when the issue is user-visible, the diff supports the root cause or repair direction, the review adds rather than restates a requirement, and the governed control flow, API, error behavior, state, or representation can be validated independently. The LLM must identify a plausible functional but non-compliant repair, justify non-implication and transferability, and propose separate executable oracles. Candidates lacking code support, separability, or distinct oracles are rejected.

Each retained candidate becomes a structured \emph{constraint seed} that preserves the normalized engineering intent and behavioral scope while removing incidental repository wording. It records the constraint, category, rationale, comments, and code evidence; the issue, root cause, relevant change, bug pattern, and an example non-compliant repair; scenario features and retrieval cues for compatible contexts; and non-redundancy, transferability, and separate oracle specifications. Complete issue, review, code, and diff provenance remains linked by a stable identifier. At this stage, oracle proposals establish suitability for instantiation; executable tests are realized only after transfer.

\subsubsection{Instance Synthesis}
Building on SWE-Mirror's demonstrated cross-repository transfer paradigm \cite{wang2025swemirrorscalingissueresolvingdatasets}, SWE-Gate transfers a constraint seed into a compatible target repository, jointly instantiating its functional bug pattern and review-derived constraint as an executable repair task with separate oracles.

The agent first explores the repository's source files, tests, existing abstractions, and implementation conventions to locate a compatible synthesis anchor. An anchor is a concrete implementation context in which the agent can introduce a user-visible functional failure while preserving an independently testable review constraint. Candidate anchors are rejected when the constraint is not naturally motivated by the target context, every plausible functional repair would necessarily satisfy the constraint, or the functional and constraint dimensions cannot be validated separately. 

Rather than generating all mutually dependent artifacts at once, the synthesis agent follows a progressive two-phase generate--execute--refine workflow. First, it creates \texttt{mutant.patch} and a functional-only \texttt{function\_test.patch}, verifying that the test passes on the original repository and fails on the mutant. It revises the anchor, injected bug, or test until this relationship holds. Second, it creates \texttt{constraint\_test.patch}, a natural functional but non-compliant \texttt{non-compliant.patch}, and a compliant \texttt{gold.patch}. The former must pass the functional test and fail the constraint test; the latter must pass both. Execution feedback refines the test and repairs. If a failure shows that the functional task cannot support a separable violation, the agent revisits the first phase, regenerates dependent artifacts, and reruns the full matrix. This progressive construction strategy first establishes a valid functional task and then adds the stricter constraint-compliance relationships, reducing the difficulty of producing all interdependent patches and tests correctly in a single generation step. 

Let $F$ and $C$ denote functional and constraint tests, and $R$, $M$, $N$, and $G$ the original repository, mutant patch, non-compliant repair, and gold repair. Valid instances satisfy Table~\ref{tab:validation_matrix}.
\begin{table}[t]
\centering
\small
\begin{tabular}{lccc}
\toprule
\textbf{Repository State} & \textbf{Notation} & \textbf{$F$} & \textbf{$C$} \\
\midrule
Original Repository & $R$ & Pass & -- \\
Mutant Bug & $R+M$ & Fail & -- \\
Non-compliant Repair & $R+M+N$ & Pass & Fail \\
Gold Repair & $R+M+G$ & Pass & Pass \\
\bottomrule
\end{tabular}
\caption{Validation matrix for a SWE-Gate instance. Following the fail-to-pass repair paradigm \cite{jimenez2024swebench}, the original repository passes $F$ and the mutant one fails it. The non-compliant repair passes $F$ but fails $C$, whereas the gold repair passes both, establishing constraint separability and satisfiability.}
\label{tab:validation_matrix}
\end{table}
Finally, the agent writes \texttt{issue.md} in a realistic user or contributor voice, describing only the visible failure without exposing injected locations, hidden tests, reference repairs, or validation internals. Any included reproduction program must run against the constructed buggy state, although this requirement is prompt-guided rather than independently rule-enforced. The separate \texttt{constraint.md} states the review-derived requirement, remains aligned with its test, and avoids revealing an expected implementation.

\subsection{Quality Assurance}

Executable validity alone does not exclude artificial, semantically inconsistent, or otherwise unsuitable LLM-generated instances. We manually inspected pilot candidates, consolidated recurring semantic rejection reasons into a failure-pattern taxonomy, and encoded this human-derived taxonomy as criteria for scalable LLM review.

Each candidate first undergoes basic structural checks to ensure that the required artifacts, execution commands, and provenance fields are complete. It is then validated in a containerized environment using the complete validation matrix: the original repository must pass the functional test, the injected bug must fail it, the non-compliant repair must pass the functional test but fail the constraint test, and the gold repair must pass both tests. Patch-application failures, execution timeouts, and other infrastructure errors are not treated as expected test failures. This Docker-based validation automatically rejects candidates whose functional and constraint behaviors cannot be independently established.

Only candidates that pass the Docker validation matrix proceed to LLM-based semantic review. The reviewer receives the constraint seed, source provenance, generated artifacts, and execution results, and applies the rejection criteria derived from manual inspection. Common semantic failure patterns include constraints that merely restate the functional issue, constraint descriptions that prescribe a particular API or implementation strategy, constraint tests that enforce behavior not stated in the natural-language description, benchmark-specific helpers or abstractions introduced solely to make the constraint testable, constraints that are not naturally motivated by the synthesized scenario, and descriptions that expose hidden evaluation artifacts. The complete taxonomy and corresponding rejection rationales are provided in the supplementary material .

Finally, all instances retained after LLM-based screening are manually inspected. This final review verifies that the issue resembles a realistic user-reported defect, the constraint remains faithful to its review-derived seed and arises naturally from the target repository, the issue and constraint are non-redundant, the non-compliant repair represents a natural functional solution, and the gold patch provides a general repository-consistent repair rather than hard-coding the generated tests. Stage-wise candidate counts and rejection reasons are reported in the supplementary material.

\subsection{Dataset Characteristics}

The current version of SWE-Gate contains 303 repository-level repair instances covering 75 open-source Python repositories. SWE-Gate covers multiple software domains, including data analysis, web frameworks, testing frameworks, command-line tools, configuration management, symbolic computation, and data validation. This repository diversity enables the benchmark to evaluate review constraints across different code contexts, rather than only measuring local conventions in a single project or a single project family.

Because a constraint may involve several engineering concerns, SWE-Gate uses a multi-label taxonomy. The most frequent categories are \emph{Error Semantics} (152 instances, 50.2\%) and \emph{Schema / Metadata / Typing} (143, 47.2\%). The taxonomy also covers ordering and argument preservation, encoding and escaping, scope generalization, compatibility, sentinel distinctions, performance, idempotence, and resource lifecycle requirements. Table~\ref{tab:constraint_analysis} reports the complete category counts together with category-level model performance. This distribution shows that code review evaluates interface behavior, exception semantics, metadata consistency, and other engineering properties beyond whether functional tests pass.

\section{Evaluation}

\subsection{Experimental Setup}

We evaluate all 303 instances. For each, an agent receives the target repository and issue description and generates a repair. Its patch is applied to a clean repository containing the injected bug and evaluated by both functional and constraint suites. Any generated changes to benchmark test files are discarded before execution, preventing evaluation leakage and ensuring that only the repair patch is assessed.

Under \emph{Constraint-Provided} ($+C$), the agent also receives the natural-language constraint; under \emph{Constraint-Omitted} ($-C$), it receives only the same repository and issue. Instances, framework, interaction budget, and hidden functional and constraint tests are identical, so the only difference is whether constraint guidance is visible during repair. The $+C$ condition measures intended SWE-Gate performance, while $-C$ is a controlled input ablation.

\paragraph{Evaluation Models.}

We evaluate GPT-5.5, GPT-5.4-mini, DeepSeek-V4-Flash, and GPT-4o-mini, spanning providers and capability levels that support repository-level engineering. All use Mini-SWE-Agent with at most 100 interaction steps. The standard SWE-Bench execution pipeline validates each patch in an isolated container.

\paragraph{Evaluation Metrics.}

We report three complementary metrics. Let $N$ be the total number of instances, $N_F$ the number of generated patches that pass the functional tests, and $N_{F\cap C}$ the number that pass both the functional and constraint tests.

\begin{itemize}
    \item \textbf{Functional Success Rate (FSR).} The proportion of all instances for which the generated patch resolves the reported functional issue:
    \[
    \mathrm{FSR}=\frac{N_F}{N}.
    \]

    \item \textbf{Constraint Following Rate (CFR).} The proportion of functionally successful repairs that also satisfy the review constraint:
    \[
    \mathrm{CFR}=\frac{N_{F\cap C}}{N_F}.
    \]

    \item \textbf{Joint Success Rate (JSR).} The proportion of all instances for which the generated patch satisfies both the functional requirement and the review constraint:
    \[
    \mathrm{JSR}=\frac{N_{F\cap C}}{N}.
    \]
\end{itemize}

JSR is the primary metric of SWE-Gate because it captures complete success under the benchmark's dual evaluation protocol. For the input ablation, we also report the percentage-point difference between the two conditions. For any metric $M\in\{\mathrm{FSR},\mathrm{CFR},\mathrm{JSR}\}$, we define

\[
\Delta M=M_{+C}-M_{-C}.
\]

A positive $\Delta M$ indicates a higher observed rate when the review constraint is provided to the agent.

\subsection{RQ1: How Do Coding Agents Perform under SWE-Gate's Dual Evaluation?}

Table~\ref{tab:overall_performance} reports functional and joint performance under the intended $+C$ setting.
\begin{table}[t]
\centering
\small
\begin{tabular*}{\columnwidth}{@{\extracolsep{\fill}}lccccc}
\toprule
\textbf{Model} &
\textbf{F.Pass} &
\textbf{J.Pass} &
\textbf{FSR} &
\textbf{CFR} &
\textbf{JSR} \\
\midrule
GPT-5.5           & 227 & 160 & \textbf{74.9} & \textbf{70.5} & \textbf{52.8} \\
GPT-5.4-mini      & 187 & 120 & 61.7 & 64.2 & 39.6 \\
DeepSeek-V4-Flash & 202 & 130 & 66.7 & 64.4 & 42.9 \\
GPT-4o-mini       &  28 &  13 &  9.2 & 46.4 &  4.3 \\
\bottomrule
\end{tabular*}
\caption{Overall performance in the Constraint-Provided setting. Rates are reported as percentages. \emph{F. Pass} denotes the number of generated patches that pass the functional test suite. 
\emph{J. Pass} (\emph{Joint Pass}) denotes the number of patches that pass both the functional and constraint test suites.}
\label{tab:overall_performance}
\end{table}

The models differ substantially in functional repair performance. GPT-5.5 achieves the highest FSR at 74.9\%, followed by DeepSeek-V4-Flash at 66.7\% and GPT-5.4-mini at 61.7\%. GPT-4o-mini resolves only 9.2\% of the instances, showing that SWE-Gate remains difficult for a substantially weaker model.

Functional success, however, does not imply joint success. GPT-5.5 produces 227 functionally successful repairs, but only 160 also pass constraint validation. The corresponding CFR is 70.5\%, meaning that 29.5\% of its functionally successful repairs violate the accompanying constraint. The same gap appears for every evaluated model.

\paragraph{Failures Hidden by Functional-Only Evaluation.}

We call a patch that passes functional validation but fails constraint validation a \emph{hidden failure}. Such a patch would be accepted under an evaluation protocol that observes only issue-related functional tests. We define the Hidden Failure Rate (HFR) as

\[
\mathrm{HFR}
=\frac{N_F-N_{F\cap C}}{N_F}
=1-\mathrm{CFR}.
\]

\begin{table}[t]
\centering
\small
\begin{tabular*}{\columnwidth}{@{\extracolsep{\fill}}lrrr}
\toprule
\textbf{Model} & \textbf{F. Pass} & \textbf{Hidden Failures} &
\textbf{HFR} \\
\midrule
GPT-5.5           & 227 & 67 & 29.5 \\
GPT-5.4-mini      & 187 & 67 & 35.8 \\
DeepSeek-V4-Flash & 202 & 72 & 35.6 \\
GPT-4o-mini       &  28 & 15 & 53.6 \\
\midrule
Total             & 644 & 221 & 34.3 \\
\bottomrule
\end{tabular*}
\caption{Functionally successful repairs that fail review constraint
validation in the Constraint-Provided setting. Rates are percentages.}
\label{tab:hidden_failures}
\end{table}

Across the four models, 644 generated patches pass the functional tests, but only 423 pass both test suites. SWE-Gate therefore identifies 221 hidden failures, corresponding to 34.3\% of all functional successes. The HFR ranges from 29.5\% for GPT-5.5 to 53.6\% for GPT-4o-mini. These results demonstrate that functional-only evaluation leaves a substantial fraction of constraint-violating repairs undetected. More precisely, they show that functional success does not entail review constraint compliance under SWE-Gate's executable evaluation protocol.

\paragraph{Answer to RQ1.}

Current coding agents resolve a substantial fraction of SWE-Gate's functional issues, but their joint success rates are consistently lower than their functional success rates. Across all evaluated model--instance pairs, 221 of 644 functional successes fail constraint validation. SWE-Gate therefore provides an additional, independently executable evaluation signal that is not captured by functional tests alone.

\subsection{RQ2: How Does Explicit Constraint Guidance Affect Repair Outcomes?  }
\begin{table*}[t]
\centering
\setlength{\tabcolsep}{4pt}
\begin{tabular}{lrrr rrr rrr}
\toprule
& \multicolumn{3}{c}{\textbf{FSR}} &
\multicolumn{3}{c}{\textbf{CFR}} &
\multicolumn{3}{c}{\textbf{JSR}} \\
\cmidrule(lr){2-4}\cmidrule(lr){5-7}\cmidrule(lr){8-10}
\textbf{Model} & $-C$ & $+C$ & $\Delta$ &
$-C$ & $+C$ & $\Delta$ &
$-C$ & $+C$ & $\Delta$ \\
\midrule
GPT-5.5           & 75.6 & 74.9 & -0.7 & 54.6 & 70.5 & +15.9 & 41.3 & 52.8 & +11.5 \\
GPT-5.4-mini      & 71.6 & 61.7 & -9.9& 50.7 & 64.2 & +13.5 & 36.3 & 39.6 &  +3.3 \\
DeepSeek-V4-Flash & 70.0 & 66.7 & -3.3 & 54.2 & 64.4 & +10.2 & 38.0 & 42.9 &  +4.9 \\
GPT-4o-mini       & 15.8 &  9.2 & -6.6 & 20.8 & 46.4 & +25.6 &  3.3 &  4.3 &  +1.0 \\
\bottomrule
\end{tabular}
\caption{Constraint input ablation. $+C$ provides the natural-language review constraint; $-C$ omits it while retaining the same hidden functional and constraint tests. $\Delta$ is $+C$ minus $-C$ in percentage points.}
\label{tab:constraint_ablation}
\end{table*}
\begin{table*}[t]
\centering
\small
\setlength{\tabcolsep}{3pt}
\begin{tabular}{l c ccc ccc ccc ccc}
\toprule
\multirow{2}{*}{Constraint Category} &
\multirow{2}{*}{N} &
\multicolumn{3}{c}{GPT-5.5} &
\multicolumn{3}{c}{GPT-5.4-mini} &
\multicolumn{3}{c}{DeepSeek-V4-Flash} &
\multicolumn{3}{c}{GPT-4o-mini} \\
\cmidrule(lr){3-5}\cmidrule(lr){6-8}\cmidrule(lr){9-11}\cmidrule(lr){12-14}
& & FSR & CFR & JSR & FSR & CFR & JSR & FSR & CFR & JSR & FSR & CFR & JSR \\
\midrule
Error semantics
&152 &75.0&72.8&54.6 &67.1&72.5&48.7 &70.4&66.4&46.7 &9.2&78.6&7.2\\
Schema / metadata / typing
&143 &72.7&68.3&49.7 &62.2&60.7&37.8 &64.3&64.1&41.3 &11.9&52.9&6.3\\
Ordering / argument preservation
&86 &60.5&76.9&46.5 &51.2&75.0&38.4 &62.8&79.6&50.0 &7.0&66.7&4.7\\
Encoding / escaping / quoting
&74 &81.1&68.3&55.4 &60.8&51.1&31.1 &79.7&55.9&44.6 &5.4&0.0&0.0\\
Scope generalization
&62 &87.1&63.0&54.8 &66.1&46.3&30.6 &69.4&53.5&37.1 &12.9&0.0&0.0\\
Compatibility / deprecation
&55 &74.5&75.6&56.4 &56.4&71.0&40.0 &63.6&77.1&49.1 &12.7&28.6&3.6\\
Missing vs. empty / sentinel distinction
&51 &74.5&81.6&60.8 &60.8&74.2&45.1 &56.9&75.9&43.1 &17.6&66.7&11.8\\
Performance / structure
&41 &75.6&71.0&53.7 &68.3&78.6&53.7 &75.6&74.2&56.1 &9.8&75.0&7.3\\
Idempotence / duplicate processing
&30 &60.0&72.2&43.3 &46.7&71.4&33.3 &46.7&71.4&33.3 &0.0&--&0.0\\
Lifecycle cleanup / resource
&19 &84.2&62.5&52.6 &68.4&53.8&36.8 &73.7&57.1&42.1 &10.5&50.0&5.3\\
\bottomrule
\end{tabular}
\caption{Performance by review constraint category in the Constraint-Provided condition. FSR and JSR are calculated over all instances in each category, whereas CFR is calculated over functionally successful repairs. ``--'' indicates that CFR is undefined because the model has no functional success in that category. Categories are multi-label and therefore not mutually exclusive.}
\label{tab:constraint_analysis}
\end{table*}

Table~\ref{tab:constraint_ablation} compares the controlled input conditions.

Providing the constraint increases JSR for every model. The largest gain is observed for GPT-5.5, whose JSR rises from 41.3\% to 52.8\%, an improvement of 11.5 percentage points. DeepSeek-V4-Flash and GPT-5.4-mini improve by 4.9 and 3.3 points, respectively, while GPT-4o-mini improves by 1.0 point. Across all four models, the number of joint successes increases from 360 to 423.

Providing the constraint description also substantially improves CFR for all four models. GPT-5.5 achieves the highest CFR improvement, increasing from 54.6\% under the Constraint-Omitted condition to 70.5\% under the Constraint-Provided condition. The CFR of GPT-5.4-mini increases from 50.7\% to 64.2\%, while that of DeepSeek-V4-Flash increases from 54.2\% to 64.4\%. GPT-4o-mini exhibits the largest relative change, with its CFR increasing from 20.8\% to 46.4\%. Overall, providing the constraint improves CFR by 10.2--25.6 percentage points across the evaluated models. These results show that, among functionally successful repairs, explicit constraint guidance substantially increases the likelihood of satisfying the corresponding engineering requirement.

At the same time, FSR does not improve under the Constraint-Provided condition. It decreases slightly for GPT-5.5 and by 3.3--9.9 percentage points for the other models. One possible explanation is that satisfying an additional requirement increases the complexity of the repair and may steer an agent away from a simpler functionally adequate patch. Because each condition contains one generation per model and instance, these results should be interpreted as an observed trade-off under the controlled input ablation rather than as a general causal claim about model behavior.

\paragraph{Answer to RQ2.}

Explicit constraint descriptions improve the observed joint success rate for all evaluated models and substantially increase the fraction of functional repairs that also satisfy constraint validation. However, this improvement is accompanied by lower functional success for three models and a small decrease for GPT-5.5. Constraint information therefore improves compliance and overall joint success in these runs, but does not uniformly improve functional repair.

\subsection{RQ3: Which Review Constraint Categories Remain Challenging?}

Table~\ref{tab:constraint_analysis} analyzes overlapping categories under $+C$; rows are not mutually exclusive and their counts should not be summed. For the three stronger models, Scope Generalization has CFRs of 63.0\%, 46.3\%, and 53.5\%, while Lifecycle Cleanup/Resource reaches 62.5\%, 53.8\%, and 57.1\%. These constraints demand coverage beyond the immediate failure or preservation across a resource lifecycle, making them difficult to satisfy with a narrowly localized patch.

Encoding/Escaping/Quoting is also difficult for GPT-5.4-mini and DeepSeek-V4-Flash, which satisfy only 51.1\% and 55.9\% of constraints after functional success, while Schema/Metadata/Typing yields 60.7--68.3\% across stronger models. By contrast, Missing-vs.-Empty/Sentinel Distinction reaches 74.2--81.6\%, and Ordering/Argument Preservation 75.0--79.6\%. The larger functional--joint gaps in the former categories therefore reflect conditional difficulty in satisfying the review constraint, not only variation in functional repair rates.

Model profiles also differ: DeepSeek-V4-Flash leads Ordering/Argument Preservation at 79.6\% CFR, GPT-5.4-mini leads Performance/Structure at 78.6\%, and GPT-5.5 reaches 81.6\% on Missing-vs.-Empty/Sentinel Distinction. GPT-4o-mini has too few functional successes in most categories for stable conclusions, so its high conditional values should not be interpreted as superior constraint following. Since categories overlap and several are small, all comparisons are descriptive, but they show that aggregate scores obscure which engineering requirements agents fail.

\paragraph{Answer to RQ3.}
Constraint-following difficulty is not uniform across engineering requirements. Among functionally successful repairs, Scope Generalization, Lifecycle Cleanup/Resource, Encoding/Escaping/Quoting, and Schema/Metadata/Typing yield some of the lowest CFR values for the three stronger models. In contrast, Missing-vs.-Empty/Sentinel Distinction and Ordering/Argument Preservation are satisfied more frequently. These results show that aggregate scores hide meaningful differences in the review constraints that agents fail to follow.

\section{Conclusion}

We introduced \textbf{SWE-Gate}, a repository-level benchmark that derives review constraints from real pull request reviews, constructs repair tasks around them, and evaluates constraint compliance separately from functional correctness. Across 303 instances from 75 Python repositories, 221 of 644 functionally successful repairs fail to satisfy the provided constraints, showing that functional-only evaluation overestimates agents' ability to satisfy complete repair requirements. Future work should extend SWE-Gate beyond Python and develop reliable evaluation methods for review requirements that cannot yet be expressed as executable tests. Incorporating broader maintainer feedback into instance construction could further improve the realism of transferred tasks and reduce potential bias introduced by LLM-assisted synthesis and screening.

\section*{Acknowledgments}

This work is supported by the National Natural Science Foundation of China (Grant No. 92582202, No. 62302534)

\bibliography{aaai2027}

@misc{shi2023sotanaopensourcesoftwaredevelopment,
      title={SoTaNa: The Open-Source Software Development Assistant}, 
      author={Ensheng Shi and Fengji Zhang and Yanlin Wang and Bei Chen and Lun Du and Hongyu Zhang and Shi Han and Dongmei Zhang and Hongbin Sun},
      year={2023},
      eprint={2308.13416},
      archivePrefix={arXiv},
      primaryClass={cs.SE},
      url={https://arxiv.org/abs/2308.13416}, 
}

@article{guo2025omnigirl,
  title={Omnigirl: A multilingual and multimodal benchmark for github issue resolution},
  author={Guo, Lianghong and Tao, Wei and Jiang, Runhan and Wang, Yanlin and Chen, Jiachi and Liu, Xilin and Ma, Yuchi and Mao, Mingzhi and Zhang, Hongyu and Zheng, Zibin},
  journal={Proceedings of the ACM on Software Engineering},
  volume={2},
  number={ISSTA},
  pages={24--46},
  year={2025},
  publisher={ACM New York, NY, USA}
}

@article{jiang2026phoenixrepair,
  title={PhoenixRepair: Rethinking Repair Strategy Exploration in Software Agents},
  author={Jiang, Tianyue and Wang, Yanlin and He, Xin and Guo, Daya and Chen, Jiachi and Wen, Ming and Shi, Ensheng and Liu, Xilin and Ma, Yuchi and Li, Guanbin},
  journal={arXiv preprint arXiv:2607.18859},
  year={2026}
}

@inproceedings{li2024repomincoder,
  title={Repomincoder: Improving repository-level code generation based on information loss screening},
  author={Li, Yifan and Shi, Ensheng and Zheng, Dewu and Duan, Kefeng and Chen, Jiachi and Wang, Yanlin},
  booktitle={Proceedings of the 15th Asia-Pacific Symposium on Internetware},
  pages={229--238},
  year={2024}
}

@inproceedings{zheng2025humanevo,
  title={Humanevo: An evolution-aware benchmark for more realistic evaluation of repository-level code generation},
  author={Zheng, Dewu and Wang, Yanlin and Shi, Ensheng and Zhang, Ruikai and Ma, Yuchi and Zhang, Hongyu and Zheng, Zibin},
  booktitle={2025 IEEE/ACM 47th International Conference on Software Engineering (ICSE)},
  pages={1372--1384},
  year={2025},
  organization={IEEE}
}

@misc{zheng2026sweprimefewertrajectoriesbetter,
      title={SWE-Prime: Fewer Trajectories, Better Performance}, 
      author={Dewu Zheng and Ruizhe Ye and Yanlin Wang and Yang Ye and Hongyu Zhang and Ensheng Shi and Xilin Liu and Yuchi Ma and Jianxing Yu and Zibin Zheng},
      year={2026},
      eprint={2608.27449},
      archivePrefix={arXiv},
      primaryClass={cs.SE},
      url={https://arxiv.org/abs/2608.27449}, 
}

@article{magis2024,
  title={Magis: Llm-based multi-agent framework for github issue resolution},
  author={Tao, Wei and Zhou, Yucheng and Wang, Yanlin and Zhang, Wenqiang and Zhang, Hongyu and Cheng, Yu},
  journal={Advances in Neural Information Processing Systems},
  volume={37},
  pages={51963--51993},
  year={2024}
}

@article{zhang2025llmhallucinationspracticalcode,
  title={Llm hallucinations in practical code generation: Phenomena, mechanism, and mitigation},
  author={Zhang, Ziyao and Wang, Chong and Wang, Yanlin and Shi, Ensheng and Ma, Yuchi and Zhong, Wanjun and Chen, Jiachi and Mao, Mingzhi and Zheng, Zibin},
  journal={Proceedings of the ACM on Software Engineering},
  volume={2},
  number={ISSTA},
  pages={481--503},
  year={2025},
  publisher={ACM New York, NY, USA}
}

@inproceedings{realsecbench2026,
  title={RealSec-bench: A benchmark for evaluating secure code generation in real-world repositories},
  author={Wang, Yanlin and Zhang, Ziyao and Wang, Chong and Xu, Xinyi and Liu, Mingwei and Wang, Yong and Chen, Jiachi and Zheng, Zibin},
  booktitle={Findings of the Association for Computational Linguistics: ACL 2026},
  pages={35866--35883},
  year={2026}
}

@article{wang2024repotransbench,
  title={RepoTransBench: A Real-World Multilingual Benchmark for Repository-Level Code Translation},
  author={Wang, Yanli and Wang, Yanlin and Wang, Suiquan and Guo, Daya and Chen, Jiachi and Grundy, John and Liu, Xilin and Ma, Yuchi and Mao, Mingzhi and Zhang, Hongyu and others},
  journal={arXiv preprint arXiv:2412.17744},
  year={2024}
}

@article{zheng2024realisticevaluation,
  title={Towards more realistic evaluation of llm-based code generation: an experimental study and beyond},
  author={Zheng, Dewu and Wang, Yanlin and Shi, Ensheng and Zhang, Ruikai and Ma, Yuchi and Zhang, Hongyu and Zheng, Zibin},
  journal={arXiv preprint arXiv:2406.06918},
  year={2024}
}

@inproceedings{arkrepobench2026,
  title={ArkRepoBench: A Repository-Level Code Completion Benchmark for HarmonyOS Development},
  author={Wang, Yanlin and Zhang, Bowen and Wang, Yanli and Guo, Daya and Zhuo, Terry Yue and Chen, Jiachi and Liu, Mingwei and Zhang, Xingong and Zheng, Zibin},
  booktitle={Findings of the Association for Computational Linguistics: ACL 2026},
  pages={19409--19429},
  year={2026}
}

@inproceedings{wang2025rlcoder,
  title={Rlcoder: Reinforcement learning for repository-level code completion},
  author={Wang, Yanlin and Wang, Yanli and Guo, Daya and Chen, Jiachi and Zhang, Ruikai and Ma, Yuchi and Zheng, Zibin},
  booktitle={2025 IEEE/ACM 47th International Conference on Software Engineering (ICSE)},
  pages={1140--1152},
  year={2025},
  organization={IEEE}
}

@article{gu2026retrievalcodegen,
  title={What to retrieve for effective retrieval-augmented code generation? an empirical study and beyond},
  author={Gu, Wenchao and Chen, Juntao and Wang, Yanlin and Jiang, Tianyue and Li, Xingzhe and Liu, Mingwei and Liu, Xilin and Ma, Yuchi and Zheng, Zibin},
  journal={arXiv preprint arXiv:2503.20589},
  year={2025}
}

@article{li2026knowledgeunittest,
  title={Knowledge Matters: Injecting Project and Testing Knowledge into LLM-based Unit Test Generation},
  author={Li, Anji and Liu, Mingwei and Chen, Zhenxi and Pei, Zheng and Li, Zike and Dai, Dekun and Wang, Yanlin and Zheng, Zibin},
  journal={arXiv preprint arXiv:2511.14224},
  year={2025}
}

@article{wang2021cocosum,
  title={Cocosum: Contextual code summarization with multi-relational graph neural network},
  author={Wang, Yanlin and Shi, Ensheng and Du, Lun and Yang, Xiaodi and Hu, Yuxuan and Han, Shi and Zhang, Hongyu and Zhang, Dongmei},
  journal={arXiv preprint arXiv:2107.01933},
  year={2021}
}

@article{wang2026contextutilization,
  title={Towards an understanding of context utilization in code intelligence},
  author={Wang, Yanlin and Duan, Kefeng and Zheng, Dewu and Shi, Ensheng and Zhang, Fengji and Wang, Yanli and Chen, Jiachi and Liu, Xilin and Ma, Yuchi and Zhang, Hongyu and others},
  journal={ACM Computing Surveys},
  volume={58},
  number={11},
  pages={1--43},
  year={2026},
  publisher={ACM New York, NY}
}

@article{guo2023snippetcomment,
  title={Snippet comment generation based on code context expansion},
  author={Guo, Hanyang and Chen, Xiangping and Huang, Yuan and Wang, Yanlin and Ding, Xi and Zheng, Zibin and Zhou, Xiaocong and Dai, Hong-Ning},
  journal={ACM Transactions on Software Engineering and Methodology},
  volume={33},
  number={1},
  pages={1--30},
  year={2023},
  publisher={ACM New York, NY}
}

@inproceedings{wang2024sparsecoder,
  title={Sparsecoder: Identifier-aware sparse transformer for file-level code summarization},
  author={Wang, Yanlin and Huang, Yanxian and Guo, Daya and Zhang, Hongyu and Zheng, Zibin},
  booktitle={2024 IEEE International Conference on Software Analysis, Evolution and Reengineering (SANER)},
  pages={614--625},
  year={2024},
  organization={IEEE}
}

@inproceedings{lin2026bettercodeunderstanding,
  title={Towards better code understanding in decoder-only models with contrastive learning},
  author={Lin, Jiayi and Wang, Yanlin and Yang, Yibiao and Zhang, Lei and Xie, Yutao},
  booktitle={Proceedings of the AAAI Conference on Artificial Intelligence},
  volume={40},
  number={38},
  pages={32006--32014},
  year={2026}
}

@inproceedings{guo2022unixcoder,
  title={Unixcoder: Unified cross-modal pre-training for code representation},
  author={Guo, Daya and Lu, Shuai and Duan, Nan and Wang, Yanlin and Zhou, Ming and Yin, Jian},
  booktitle={Proceedings of the 60th Annual Meeting of the Association for Computational Linguistics (Volume 1: Long Papers)},
  pages={7212--7225},
  year={2022}
}

@article{Guo2025SWEDC,
  title={SWE Data Construction, Automatically!},
  author={Lianghong Guo and Yanlin Wang and Caihua Li and Wei Tao and Pengyu Yang and Jiachi Chen and Haoyu Song and Duyu Tang and Zibin Zheng},
  journal={Proceedings of the ACM on Software Engineering},
  year={2025},
  volume={3},
  pages={525 - 546},
  url={https://api.semanticscholar.org/CorpusID:284488853},
  note={\textcolor{red}{\textsuperscript{*}}},
}

@misc{zheng2025generalperformancedomain,
      title={Top General Performance = Top Domain Performance? DomainCodeBench: A Multi-domain Code Generation Benchmark}, 
      author={Dewu Zheng and Yanlin Wang and Ensheng Shi and Xilin Liu and Yuchi Ma and Hongyu Zhang and Zibin Zheng},
      year={2025},
      eprint={2412.18573},
      archivePrefix={arXiv},
      primaryClass={cs.SE},
      url={https://arxiv.org/abs/2412.18573}, 
}

@misc{jiang2026aligncoderaligningretrievaltarget,
      title={AlignCoder: Aligning Retrieval with Target Intent for Repository-Level Code Completion}, 
      author={Tianyue Jiang and Yanli Wang and Yanlin Wang and Daya Guo and Ensheng Shi and Yuchi Ma and Jiachi Chen and Zibin Zheng},
      year={2026},
      eprint={2601.19697},
      archivePrefix={arXiv},
      primaryClass={cs.SE},
      url={https://arxiv.org/abs/2601.19697}, 
}

@article{wang2026reporeasoner,
  title={RepoReasoner: Evaluating Repository-Level Code Reasoning Ability of Long-Context Language Models},
  author={Wang, Yanlin and Wang, Suiquan and Wang, Yanli and Zhang, Bowen and Guo, Daya and Chen, Jiachi and Zheng, Zibin},
  journal={Proceedings of the ACM on Software Engineering},
  volume={3},
  number={FSE},
  pages={2790--2812},
  year={2026},
  publisher={ACM New York, NY, USA}
}

@INPROCEEDINGS{LargeLanguageModelsAreQualifiedBenchmarkBuilders,
  author={Yang, Kang and Mao, Xinjun and Wang, Shangwen and Wang, Yanlin and Zhang, Tanghaoran and Lin, Bo and Qin, Yihao and Zhang, Zhang and Lu, Yao and Al-Sabahi, Kamal},
  booktitle={2025 IEEE/ACM 33rd International Conference on Program Comprehension (ICPC)}, 
  title={Large Language Models Are Qualified Benchmark Builders: Rebuilding Pre-Training Datasets for Advancing Code Intelligence Tasks}, 
  year={2025},
  volume={},
  number={},
  pages={298-309},
  doi={10.1109/ICPC66645.2025.00038}}

@misc{wang2024agentssoftwareengineeringsurvey,
      title={Agents in Software Engineering: Survey, Landscape, and Vision}, 
      author={Yanlin Wang and Wanjun Zhong and Yanxian Huang and Ensheng Shi and Min Yang and Jiachi Chen and Hui Li and Yuchi Ma and Qianxiang Wang and Zibin Zheng},
      year={2024},
      eprint={2409.09030},
      archivePrefix={arXiv},
      primaryClass={cs.SE},
      url={https://arxiv.org/abs/2409.09030}, 
}

@misc{zheng2024understandinglargelanguagemodels,
      title={Towards an Understanding of Large Language Models in Software Engineering Tasks}, 
      author={Zibin Zheng and Kaiwen Ning and Qingyuan Zhong and Jiachi Chen and Wenqing Chen and Lianghong Guo and Weicheng Wang and Yanlin Wang},
      year={2024},
      eprint={2308.11396},
      archivePrefix={arXiv},
      primaryClass={cs.SE},
      url={https://arxiv.org/abs/2308.11396}, 
}

@misc{zheng2024surveylargelanguagemodels,
      title={A Survey of Large Language Models for Code: Evolution, Benchmarking, and Future Trends}, 
      author={Zibin Zheng and Kaiwen Ning and Yanlin Wang and Jingwen Zhang and Dewu Zheng and Mingxi Ye and Jiachi Chen},
      year={2024},
      eprint={2311.10372},
      archivePrefix={arXiv},
      primaryClass={cs.SE},
      url={https://arxiv.org/abs/2311.10372}, 
}

@inproceedings{wang-etal-2024-large-language-models-fair,
    title = "Large Language Models are not Fair Evaluators",
    author = "Wang, Peiyi  and
      Li, Lei  and
      Chen, Liang  and
      Cai, Zefan  and
      Zhu, Dawei  and
      Lin, Binghuai  and
      Cao, Yunbo  and
      Kong, Lingpeng  and
      Liu, Qi  and
      Liu, Tianyu  and
      Sui, Zhifang",
    editor = "Ku, Lun-Wei  and
      Martins, Andre  and
      Srikumar, Vivek",
    booktitle = "Proceedings of the 62nd Annual Meeting of the Association for Computational Linguistics (Volume 1: Long Papers)",
    month = aug,
    year = "2024",
    address = "Bangkok, Thailand",
    publisher = "Association for Computational Linguistics",
    url = "https://aclanthology.org/2024.acl-long.511/",
    doi = "10.18653/v1/2024.acl-long.511",
    pages = "9440--9450"
}

@inproceedings{10.1145/2597073.2597082,
author = {Beller, Moritz and Bacchelli, Alberto and Zaidman, Andy and Juergens, Elmar},
title = {Modern code reviews in open-source projects: which problems do they fix?},
year = {2014},
isbn = {9781450328630},
publisher = {Association for Computing Machinery},
address = {New York, NY, USA},
url = {https://doi.org/10.1145/2597073.2597082},
doi = {10.1145/2597073.2597082},
booktitle = {Proceedings of the 11th Working Conference on Mining Software Repositories},
pages = {202–211},
numpages = {10},
location = {Hyderabad, India},
series = {MSR 2014}
}

@inproceedings{10.1145/3106237.3106274,
author = {Yang, Jinqiu and Zhikhartsev, Alexey and Liu, Yuefei and Tan, Lin},
title = {Better test cases for better automated program repair},
year = {2017},
isbn = {9781450351058},
publisher = {Association for Computing Machinery},
address = {New York, NY, USA},
url = {https://doi.org/10.1145/3106237.3106274},
doi = {10.1145/3106237.3106274},
booktitle = {Proceedings of the 2017 11th Joint Meeting on Foundations of Software Engineering},
pages = {831–841},
numpages = {11},
location = {Paderborn, Germany},
series = {ESEC/FSE 2017}
}

@inproceedings{10.1145/2786805.2786825,
author = {Smith, Edward K. and Barr, Earl T. and Le Goues, Claire and Brun, Yuriy},
title = {Is the cure worse than the disease? overfitting in automated program repair},
year = {2015},
isbn = {9781450336758},
publisher = {Association for Computing Machinery},
address = {New York, NY, USA},
url = {https://doi.org/10.1145/2786805.2786825},
doi = {10.1145/2786805.2786825},
booktitle = {Proceedings of the 2015 10th Joint Meeting on Foundations of Software Engineering},
pages = {532–543},
numpages = {12},
location = {Bergamo, Italy},
series = {ESEC/FSE 2015}
}

@inproceedings{NEURIPS2025_8b86cf5a,
 author = {Yang, John and Lieret, Kilian and Jimenez, Carlos and Wettig, Alexander and Khandpur, Kabir and Zhang, Yanzhe and Hui, Binyuan and Press, Ofir and Schmidt, Ludwig and Yang, Diyi},
 booktitle = {Advances in Neural Information Processing Systems},
 editor = {D. Belgrave and C. Zhang and H. Lin and R. Pascanu and P. Koniusz and M. Ghassemi and N. Chen},
 pages = {},
 publisher = {Curran Associates, Inc.},
 title = {SWE-smith: Scaling Data for Software Engineering Agents},
 url = {https://proceedings.neurips.cc/paper_files/paper/2025/file/8b86cf5ace600c48fd188efbb8dedec8-Paper-Datasets_and_Benchmarks_Track.pdf},
 volume = {38},
 year = {2025}
}

@inproceedings{NEURIPS2025_21bec6ac,
 author = {Badertdinov, Ibragim and Golubev, Alexander and Nekrashevich, Maksim and Shevtsov, Anton and Karasik, Simon and Andriushchenko, Andrei and Trofimova, Maria and Litvintseva, Daria and Yangel, Boris},
 booktitle = {Advances in Neural Information Processing Systems},
 editor = {D. Belgrave and C. Zhang and H. Lin and R. Pascanu and P. Koniusz and M. Ghassemi and N. Chen},
 pages = {},
 publisher = {Curran Associates, Inc.},
 title = {SWE-rebench: An Automated Pipeline for Task Collection and Decontaminated Evaluation of Software Engineering Agents},
 url = {https://proceedings.neurips.cc/paper_files/paper/2025/file/21bec6ace947b1b58967b945c8ac0f10-Paper-Datasets_and_Benchmarks_Track.pdf},
 volume = {38},
 year = {2025}
}

@inproceedings{NEURIPS2025_d83c4a74,
 author = {Zhang, Linghao and He, Shilin and Zhang, Chaoyun and Kang, Yu and Li, Bowen and Xie, Chengxing and Wang, Junhao and Wang, Maoquan and Huang, Yufan and Fu, Shengyu and Nallipogu, Elsie and Lin, Qingwei and Dang, Yingnong and Rajmohan, Saravan and Zhang, Dongmei},
 booktitle = {Advances in Neural Information Processing Systems},
 editor = {D. Belgrave and C. Zhang and H. Lin and R. Pascanu and P. Koniusz and M. Ghassemi and N. Chen},
 pages = {},
 publisher = {Curran Associates, Inc.},
 title = {SWE-bench Goes Live!},
 url = {https://proceedings.neurips.cc/paper_files/paper/2025/file/d83c4a745789690f82e86d0ef752ae7c-Paper-Datasets_and_Benchmarks_Track.pdf},
 volume = {38},
 year = {2025}
}

@inproceedings{10.1145/3368089.3417943,
author = {Widyasari, Ratnadira and Sim, Sheng Qin and Lok, Camellia and Qi, Haodi and Phan, Jack and Tay, Qijin and Tan, Constance and Wee, Fiona and Tan, Jodie Ethelda and Yieh, Yuheng and Goh, Brian and Thung, Ferdian and Kang, Hong Jin and Hoang, Thong and Lo, David and Ouh, Eng Lieh},
title = {BugsInPy: a database of existing bugs in Python programs to enable controlled testing and debugging studies},
year = {2020},
isbn = {9781450370431},
publisher = {Association for Computing Machinery},
address = {New York, NY, USA},
url = {https://doi.org/10.1145/3368089.3417943},
doi = {10.1145/3368089.3417943},
booktitle = {Proceedings of the 28th ACM Joint Meeting on European Software Engineering Conference and Symposium on the Foundations of Software Engineering},
pages = {1556–1560},
numpages = {5},
location = {Virtual Event, USA},
series = {ESEC/FSE 2020}
}

@inproceedings{10.1145/2610384.2628055,
author = {Just, Ren\'{e} and Jalali, Darioush and Ernst, Michael D.},
title = {Defects4J: a database of existing faults to enable controlled testing studies for Java programs},
year = {2014},
isbn = {9781450326452},
publisher = {Association for Computing Machinery},
address = {New York, NY, USA},
url = {https://doi.org/10.1145/2610384.2628055},
doi = {10.1145/2610384.2628055},
booktitle = {Proceedings of the 2014 International Symposium on Software Testing and Analysis},
pages = {437–440},
numpages = {4},
location = {San Jose, CA, USA},
series = {ISSTA 2014}
}

@inproceedings{
ding2023crosscodeeval,
title={CrossCodeEval: A Diverse and Multilingual Benchmark for Cross-File Code Completion},
author={Yangruibo Ding and Zijian Wang and Wasi Uddin Ahmad and Hantian Ding and Ming Tan and Nihal Jain and Murali Krishna Ramanathan and Ramesh Nallapati and Parminder Bhatia and Dan Roth and Bing Xiang},
booktitle={Thirty-seventh Conference on Neural Information Processing Systems Datasets and Benchmarks Track},
year={2023},
url={https://openreview.net/forum?id=wgDcbBMSfh}
}

@misc{wang2025swemirrorscalingissueresolvingdatasets,
      title={SWE-Mirror: Scaling Issue-Resolving Datasets by Mirroring Issues Across Repositories}, 
      author={Junhao Wang and Daoguang Zan and Shulin Xin and Siyao Liu and Yurong Wu and Kai Shen},
      year={2025},
      eprint={2509.08724},
      archivePrefix={arXiv},
      primaryClass={cs.SE},
      url={https://arxiv.org/abs/2509.08724}, 
}

@inproceedings{zhang-etal-2023-repocoder,
    title = "{R}epo{C}oder: Repository-Level Code Completion Through Iterative Retrieval and Generation",
    author = "Zhang, Fengji  and
      Chen, Bei  and
      Zhang, Yue  and
      Keung, Jacky  and
      Liu, Jin  and
      Zan, Daoguang  and
      Mao, Yi  and
      Lou, Jian-Guang  and
      Chen, Weizhu",
    editor = "Bouamor, Houda  and
      Pino, Juan  and
      Bali, Kalika",
    booktitle = "Proceedings of the 2023 Conference on Empirical Methods in Natural Language Processing",
    month = dec,
    year = "2023",
    address = "Singapore",
    publisher = "Association for Computational Linguistics",
    url = "https://aclanthology.org/2023.emnlp-main.151/",
    doi = "10.18653/v1/2023.emnlp-main.151",
    pages = "2471--2484"
}

@misc{yang2024swebenchmultimodalaisystems,
      title={SWE-bench Multimodal: Do AI Systems Generalize to Visual Software Domains?}, 
      author={John Yang and Carlos E. Jimenez and Alex L. Zhang and Kilian Lieret and Joyce Yang and Xindi Wu and Ori Press and Niklas Muennighoff and Gabriel Synnaeve and Karthik R. Narasimhan and Diyi Yang and Sida I. Wang and Ofir Press},
      year={2024},
      eprint={2410.03859},
      archivePrefix={arXiv},
      primaryClass={cs.CL},
      url={https://arxiv.org/abs/2410.03859}, 
}

@misc{zan2024swebenchjavagithubissueresolving,
      title={SWE-bench-java: A GitHub Issue Resolving Benchmark for Java}, 
      author={Daoguang Zan and Zhirong Huang and Ailun Yu and Shaoxin Lin and Yifan Shi and Wei Liu and Dong Chen and Zongshuai Qi and Hao Yu and Lei Yu and Dezhi Ran and Muhan Zeng and Bo Shen and Pan Bian and Guangtai Liang and Bei Guan and Pengjie Huang and Tao Xie and Yongji Wang and Qianxiang Wang},
      year={2024},
      eprint={2408.14354},
      archivePrefix={arXiv},
      primaryClass={cs.SE},
      url={https://arxiv.org/abs/2408.14354}, 
}

@inproceedings{NEURIPS2023_91f18a12,
 author = {Zheng, Lianmin and Chiang, Wei-Lin and Sheng, Ying and Zhuang, Siyuan and Wu, Zhanghao and Zhuang, Yonghao and Lin, Zi and Li, Zhuohan and Li, Dacheng and Xing, Eric and Zhang, Hao and Gonzalez, Joseph and Stoica, Ion},
 booktitle = {Advances in Neural Information Processing Systems},
 doi = {10.52202/075280-2020},
 editor = {A. Oh and T. Naumann and A. Globerson and K. Saenko and M. Hardt and S. Levine},
 pages = {46595--46623},
 publisher = {Curran Associates, Inc.},
 title = {Judging LLM-as-a-Judge with MT-Bench and Chatbot Arena},
 url = {https://proceedings.neurips.cc/paper_files/paper/2023/file/91f18a1287b398d378ef22505bf41832-Paper-Datasets_and_Benchmarks.pdf},
 volume = {36},
 year = {2023}
}

@misc{joshi2025swebenchclcontinuallearningcoding,
      title={SWE-Bench-CL: Continual Learning for Coding Agents}, 
      author={Thomas Joshi and Shayan Chowdhury and Fatih Uysal},
      year={2025},
      eprint={2507.00014},
      archivePrefix={arXiv},
      primaryClass={cs.LG},
      url={https://arxiv.org/abs/2507.00014}, 
}

@misc{deng2025swebenchproaiagents,
      title={SWE-Bench Pro: Can AI Agents Solve Long-Horizon Software Engineering Tasks?}, 
      author={Xiang Deng and Jeff Da and Edwin Pan and Yannis Yiming He and Charles Ide and Kanak Garg and Niklas Lauffer and Andrew Park and Nitin Pasari and Chetan Rane and Karmini Sampath and Maya Krishnan and Srivatsa Kundurthy and Sean Hendryx and Zifan Wang and Vijay Bharadwaj and Jeff Holm and Raja Aluri and Chen Bo Calvin Zhang and Noah Jacobson and Bing Liu and Brad Kenstler},
      year={2025},
      eprint={2509.16941},
      archivePrefix={arXiv},
      primaryClass={cs.SE},
      url={https://arxiv.org/abs/2509.16941}, 
}

@inproceedings{NEURIPS2025_5afa9cb1,
 author = {Zan, Daoguang and Huang, Zhirong and Liu, Wei and Chen, Hanwu and Xin, Shulin and Zhang, Linhao and Liu, Qi and Aoyan, Li and Chen, Lu and Zhong, Xiaojian and Liu, Siyao and Xiao, Yongsheng and Chen, Liangqiang and Zhang, Yuyu and Su, Jing and Liu, Tianyu and LONG, RUI and Ding, Ming and xiang, liang},
 booktitle = {Advances in Neural Information Processing Systems},
 editor = {D. Belgrave and C. Zhang and H. Lin and R. Pascanu and P. Koniusz and M. Ghassemi and N. Chen},
 pages = {},
 publisher = {Curran Associates, Inc.},
 title = {Multi-SWE-bench: A Multilingual Benchmark for Issue Resolving},
 url = {https://proceedings.neurips.cc/paper_files/paper/2025/file/5afa9cb1e917b898ad418216dc726fbd-Paper-Datasets_and_Benchmarks_Track.pdf},
 volume = {38},
 year = {2025}
}

@misc{khan2023xcodeevallargescalemultilingual,
      title={xCodeEval: A Large Scale Multilingual Multitask Benchmark for Code Understanding, Generation, Translation and Retrieval}, 
      author={Mohammad Abdullah Matin Khan and M Saiful Bari and Xuan Long Do and Weishi Wang and Md Rizwan Parvez and Shafiq Joty},
      year={2023},
      eprint={2303.03004},
      archivePrefix={arXiv},
      primaryClass={cs.CL},
      url={https://arxiv.org/abs/2303.03004}, 
}

@misc{liu2023repobenchbenchmarkingrepositorylevelcode,
      title={RepoBench: Benchmarking Repository-Level Code Auto-Completion Systems}, 
      author={Tianyang Liu and Canwen Xu and Julian McAuley},
      year={2023},
      eprint={2306.03091},
      archivePrefix={arXiv},
      primaryClass={cs.CL},
      url={https://arxiv.org/abs/2306.03091}, 
}

@misc{austin2021programsynthesislargelanguage,
      title={Program Synthesis with Large Language Models}, 
      author={Jacob Austin and Augustus Odena and Maxwell Nye and Maarten Bosma and Henryk Michalewski and David Dohan and Ellen Jiang and Carrie Cai and Michael Terry and Quoc Le and Charles Sutton},
      year={2021},
      eprint={2108.07732},
      archivePrefix={arXiv},
      primaryClass={cs.PL},
      url={https://arxiv.org/abs/2108.07732}, 
}

@misc{chen2021evaluatinglargelanguagemodels,
      title={Evaluating Large Language Models Trained on Code}, 
      author={Mark Chen and Jerry Tworek and Heewoo Jun and Qiming Yuan and Henrique Ponde de Oliveira Pinto and Jared Kaplan and Harri Edwards and Yuri Burda and Nicholas Joseph and Greg Brockman and Alex Ray and Raul Puri and Gretchen Krueger and Michael Petrov and Heidy Khlaaf and Girish Sastry and Pamela Mishkin and Brooke Chan and Scott Gray and Nick Ryder and Mikhail Pavlov and Alethea Power and Lukasz Kaiser and Mohammad Bavarian and Clemens Winter and Philippe Tillet and Felipe Petroski Such and Dave Cummings and Matthias Plappert and Fotios Chantzis and Elizabeth Barnes and Ariel Herbert-Voss and William Hebgen Guss and Alex Nichol and Alex Paino and Nikolas Tezak and Jie Tang and Igor Babuschkin and Suchir Balaji and Shantanu Jain and William Saunders and Christopher Hesse and Andrew N. Carr and Jan Leike and Josh Achiam and Vedant Misra and Evan Morikawa and Alec Radford and Matthew Knight and Miles Brundage and Mira Murati and Katie Mayer and Peter Welinder and Bob McGrew and Dario Amodei and Sam McCandlish and Ilya Sutskever and Wojciech Zaremba},
      year={2021},
      eprint={2107.03374},
      archivePrefix={arXiv},
      primaryClass={cs.LG},
      url={https://arxiv.org/abs/2107.03374}, 
}

@misc{duan2025hierarchicalevolvablebenchmarkfinegrained,
      title={A Hierarchical and Evolvable Benchmark for Fine-Grained Code Instruction Following with Multi-Turn Feedback}, 
      author={Guoliang Duan and Mingwei Liu and Yanlin Wang and Chong Wang and Xin Peng and Zibin Zheng},
      year={2025},
      eprint={2507.00699},
      archivePrefix={arXiv},
      primaryClass={cs.SE},
      url={https://arxiv.org/abs/2507.00699}, 
}

@article{Yu2026DoesPR,
  title={Does Pass Rate Tell the Whole Story? Evaluating Design Constraint Compliance in LLM-based Issue Resolution},
  author={Kai Yu and Zhenhao Zhou and Junhao Zeng and Ying Wang and Xueying Du and Zhiqiang Yuan and Junwei Liu and Ziyu Zhou and Yujia Wang and Chong Wang and Xin Peng},
  journal={ArXiv},
  year={2026},
  volume={abs/2604.05955},
  url={https://api.semanticscholar.org/CorpusID:287209367}
}

@inproceedings{10.1145/2491411.2491444,
author = {Rigby, Peter C. and Bird, Christian},
title = {Convergent contemporary software peer review practices},
year = {2013},
isbn = {9781450322379},
publisher = {Association for Computing Machinery},
address = {New York, NY, USA},
url = {https://doi.org/10.1145/2491411.2491444},
doi = {10.1145/2491411.2491444},
booktitle = {Proceedings of the 2013 9th Joint Meeting on Foundations of Software Engineering},
pages = {202–212},
numpages = {11},
location = {Saint Petersburg, Russia},
series = {ESEC/FSE 2013}
}

@inproceedings{10.1145/3183519.3183525,
author = {Sadowski, Caitlin and S\"{o}derberg, Emma and Church, Luke and Sipko, Michal and Bacchelli, Alberto},
title = {Modern code review: a case study at google},
year = {2018},
isbn = {9781450356596},
publisher = {Association for Computing Machinery},
address = {New York, NY, USA},
url = {https://doi.org/10.1145/3183519.3183525},
doi = {10.1145/3183519.3183525},
booktitle = {Proceedings of the 40th International Conference on Software Engineering: Software Engineering in Practice},
pages = {181–190},
numpages = {10},
location = {Gothenburg, Sweden},
series = {ICSE-SEIP '18}
}

@INPROCEEDINGS{6606617,
  author={Bacchelli, Alberto and Bird, Christian},
  booktitle={2013 35th International Conference on Software Engineering (ICSE)}, 
  title={Expectations, outcomes, and challenges of modern code review}, 
  year={2013},
  volume={},
  number={},
  pages={712-721},
  doi={10.1109/ICSE.2013.6606617}}

@misc{xia2024agentlessdemystifyingllmbasedsoftware,
      title={Agentless: Demystifying LLM-based Software Engineering Agents}, 
      author={Chunqiu Steven Xia and Yinlin Deng and Soren Dunn and Lingming Zhang},
      year={2024},
      eprint={2407.01489},
      archivePrefix={arXiv},
      primaryClass={cs.SE},
      url={https://arxiv.org/abs/2407.01489}, 
}

@inproceedings{
wang2025openhands,
title={OpenHands: An Open Platform for {AI} Software Developers as Generalist Agents},
author={Xingyao Wang and Boxuan Li and Yufan Song and Frank F. Xu and Xiangru Tang and Mingchen Zhuge and Jiayi Pan and Yueqi Song and Bowen Li and Jaskirat Singh and Hoang H. Tran and Fuqiang Li and Ren Ma and Mingzhang Zheng and Bill Qian and Daniel Shao and Niklas Muennighoff and Yizhe Zhang and Binyuan Hui and Junyang Lin and Robert Brennan and Hao Peng and Heng Ji and Graham Neubig},
booktitle={The Thirteenth International Conference on Learning Representations},
year={2025},
url={https://openreview.net/forum?id=OJd3ayDDoF}
}

@inproceedings{
jimenez2024swebench,
title={{SWE}-bench: Can Language Models Resolve Real-world Github Issues?},
author={Carlos E Jimenez and John Yang and Alexander Wettig and Shunyu Yao and Kexin Pei and Ofir Press and Karthik R Narasimhan},
booktitle={The Twelfth International Conference on Learning Representations},
year={2024},
url={https://openreview.net/forum?id=VTF8yNQM66}
}

@inproceedings{NEURIPS2024_5a7c9475,
 author = {Yang, John and Jimenez, Carlos and Wettig, Alexander and Lieret, Kilian and Yao, Shunyu and Narasimhan, Karthik and Press, Ofir},
 booktitle = {Advances in Neural Information Processing Systems},
 doi = {10.52202/079017-1601},
 editor = {A. Globerson and L. Mackey and D. Belgrave and A. Fan and U. Paquet and J. Tomczak and C. Zhang},
 pages = {50528--50652},
 publisher = {Curran Associates, Inc.},
 title = {SWE-agent: Agent-Computer Interfaces Enable Automated Software Engineering},
 url = {https://proceedings.neurips.cc/paper_files/paper/2024/file/5a7c947568c1b1328ccc5230172e1e7c-Paper-Conference.pdf},
 volume = {37},
 year = {2024}
}

@inproceedings{yu-etal-2025-utboost,
    title = "{UTB}oost: Rigorous Evaluation of Coding Agents on {SWE}-Bench",
    author = "Yu, Boxi  and
      Zhu, Yuxuan  and
      He, Pinjia  and
      Kang, Daniel",
    editor = "Che, Wanxiang  and
      Nabende, Joyce  and
      Shutova, Ekaterina  and
      Pilehvar, Mohammad Taher",
    booktitle = "Proceedings of the 63rd Annual Meeting of the Association for Computational Linguistics (Volume 1: Long Papers)",
    month = jul,
    year = "2025",
    address = "Vienna, Austria",
    publisher = "Association for Computational Linguistics",
    url = "https://aclanthology.org/2025.acl-long.189/",
    doi = "10.18653/v1/2025.acl-long.189",
    pages = "3762--3774",
    ISBN = "979-8-89176-251-0"
}

@inproceedings{10.1145/2771783.2771791,
author = {Qi, Zichao and Long, Fan and Achour, Sara and Rinard, Martin},
title = {An analysis of patch plausibility and correctness for generate-and-validate patch generation systems},
year = {2015},
isbn = {9781450336208},
publisher = {Association for Computing Machinery},
address = {New York, NY, USA},
url = {https://doi.org/10.1145/2771783.2771791},
doi = {10.1145/2771783.2771791},
booktitle = {Proceedings of the 2015 International Symposium on Software Testing and Analysis},
pages = {24–36},
numpages = {13},
location = {Baltimore, MD, USA},
series = {ISSTA 2015}
}

\end{document}